%% file: Simpsonetal.tex
\documentclass[sts]{imsart}

\usepackage{algorithm,algorithmic,amsfonts,amssymb,amsmath,amscd,amsthm,bm,latexsym}
\usepackage{natbib}
\usepackage{nicefrac}

\renewcommand{\rho}{\varrho}

\begin{document}

\begin{frontmatter}

\title{Some comments about ``Penalising model component complexity: A principled, practical approach to
constructing priors" by Simpson, Rue, Martins, Riebler, and S{\o}rbye}
\runtitle{Discussion of Penalising model component complexity}
\thankstext{T1}{Christian P. Robert (corresponding author)
and Judith Rousseau, CEREMADE, Universit{\' e} Paris-Dauphine, PSL, 75775 Paris cedex 16, France
{\sf \{xian,rousseau\}@ceremade.dauphine.fr}. 
Both authors are members of Laboratoire de Statistique, CREST, Paris. C.P.~Robert is also affiliated as a part-time
professor in the Department of Statistics of the University of Warwick, Coventry, UK, and is a senior member of IUF for
2016-2021.}

\begin{aug}
\author{\snm{Christian P.~Robert and Judith Rousseau}}
\affiliation{Universit{\'e} Paris-Dauphine, PSL, CNRS, CEREMADE, and CREST, Paris}
\end{aug}

\begin{abstract}
This note discusses the paper ``Penalising model component complexity" by Simpson et al. (2017). While we acknowledge the highly
novel approach to prior construction and commend the authors for setting new -encompassing principles that will Bayesian
modelling, and while we perceive the potential connection with other branches of the literature, we remain uncertain as 
to what extent the principles exposed in the paper can be developed outside specific models, given their lack of
precision. The very notions of model component, base model, overfitting prior are for instance conceptual rather than
mathematical and we thus fear the concept of penalised complexity may not further than extending first-guess priors into
larger families, thus failing to establish reference priors on a novel sound ground.
\end{abstract}

\begin{keyword}
\kwd{decision-theory}
\kwd{Gamma-minimaxity}
\kwd{misspecification}
\kwd{prior selection}
\kwd{robust methodology}
\end{keyword}
\end{frontmatter}

\section{Introduction}

\begin{quote}``On the other end of the hunt for the holy grail, ``objective'' priors are data-dependent and are not
uniformly accepted among Bayesians on philosophical grounds.''\end{quote}

The most sensitive aspect of Bayesian modelling is undoubtedly the call to a prior distribution. From Fisher onwards
\citep{zabell:1992}, up to this very day \citep{martin:liu:2015,hd}, the concept of prior distribution has been criticised
as being alien to the sampling model and critics have pointed out the arbitrariness of some or all aspects of chosen
priors. This is most prominent in weakly informative settings when the context is deemed too poor to return an expert
opinion and thus build an informed prior. The whole branch of so-called objective (aka, reference or non-informative)
Bayesian statistics \citep{berger:bernardo:sun:2009} has been constructed to answer and bypass such criticisms, clearly
not achieving a complete silencing of such criticisms.

\begin{quote}``Prior selection is the fundamental issue in Bayesian statistics. Priors are the Bayesian’s greatest tool,
but they are also the greatest point for criticism: the arbitrariness of prior selection procedures and the lack of
realistic sensitivity analysis (...) are a serious argument against current Bayesian practice.''\end{quote}

In this paper, the authors aim at providing some form of prior robust modelling, rather than non-informative principles
that are so delicate to specify, as shown by the literature \citep{liseo:2006}. It is a highly timely and pertinent paper
on the selection and construction of
priors. It also shows that the field of ``objective'' Bayes theory is still central to Bayesian statistics 
and this makes a great argument to encourage more Bayesian researchers to consider this branch of our field.
This attempt is most commendable and we hope it will induce others to enlarge and deepen the work in this direction. 

The paper starts with a review of prior selection in connection with levels of prior information. The authors then
advance some desirable principles for the construction of priors on a collection of models that is restricted to
hierarchical additive models with a latent structure. Connections with other approaches abound, from Jeffreys priors and the
asymptotic developments of \cite{bochkina:green:2014}, to the non-local priors of \cite{johnson:rossell:2010}, and
sparsity priors. (Although this may constitute the more tentative part of the paper.) The applications are the disease mapping
model of Besag et al. (1991) and the multivariate probit model.


\section{PC priors}

\begin{quote}``Most model components can be naturally regarded as a flexible version of a base model.''\end{quote}

The starting point for the authors' modelling is the so-called {\em base model}. We understand this approach operates
via the (specialised?) notion of model
components, as modularity is obviously essential (if challenging) in devising reference or default priors. However, the
obvious question it induces is to figure how easy it is to define this base model. For instance, the authors later make
a connection between base models and hypothesis testing. One may wonder whether or not such a notion always translates
into a null hypothesis formulation and whether or not  this reformulation is pertinent (if only because it relates to
tests). From a more global perspective, we remain rather skeptical that there
could be an automated version of the derivation of a base model, just as there is no single version of an ``objective" Bayes prior
\citep{kass:wasserman:1996,robert:2001}. We assume this derivation somewhat follows from the ``block'' idea but we
wonder at how generic model construction by blocks can be. The authors do acknowledge the difficulty in Section 7, as an
unrealistic expectation on the practitioners. 

In particular defining a base model is typically done under a given parameterisation, which implies the whole approach
fails to stay invariant under reparameterisation. To illustrate our point consider the discussion in Section 4.5 on
sparsity. The authors consider the model
$$ 
\mathbf y = (y_1 , \cdots , y_n) \sim \pi(\mathbf y|\beta) , \quad \beta \sim \mathcal{N}_p(0, D),
\quad D = \mbox{diag}(D_{11}, \cdots, D_{pp})\,.
$$
They define as their base model a sparse model, in essence corresponding to $\beta = 0$, that they translate into
saying $D_{ii}= 0$ for all $i$ and then into considering as a prior on $D$
\begin{equation} \label{iid}
D_{ii}^{-1} \sim \pi_\tau
\end{equation}
where $\pi_{\tau}$ is the prior defined in (3.3). As the authors note, this does not lead to a correct sparse behaviour
(in other words to convincing shrinkage) and they suggest to define as a sparse approach to the PC-prior a hierarchical
prior similar in spirit to a spike and slab, where one first selects the number of non-zero components and then for
these components only assumes \eqref{iid}.  But it seems to us wrong to assume that the original problem translates into
imposing that the $D_{ii}$'s have to be i.i.d \textit{sparse}. A more effective approach would have been to consider 
$$ 
\mathbf y = (y_1 , \cdots , y_n) \sim \pi(\mathbf y|\beta) , \quad \beta \sim \mathcal{N}_p(0, \tau^{-1} D), \quad D =
\mbox{diag}(D_{11}, \cdots, D_{pp})\,,
$$
where sparsity is expressed through $\tau \sim \pi_\tau$, as it then becomes a global notion. This modelling would have
been much closer in spirit to the horseshoe prior approach, but with a different prior on $\tau $ implied by the PC approach and, being
univariate, less problematic in the context of PC priors. 

\begin{quote} ``Occam's razor is the principle of parsimony, for which simpler model formulations should be preferred until there
is enough support for a more complex model.''\end{quote}

Assuming a base model has been constructed, we are definitely supportive of the idea of putting a prior on the distance
from the base! Even more because this concept is parameterisation invariant, at least at the hyperparameter level. And
because it somewhat gives a definitive meaning to the repeatedly invoked Occam's razor, even though we feel we could
easily live without this constant reference to a vague notion proposed by a medieval monk in England. However, unless
the hyperparameter $\xi$ is one-dimensional, this approach fails to define a prior on $\xi$ {\em per se}, which implies
making further choices for the reference prior modelling. 

We still wonder as to the particular Kullback-Leibler divergence chosen by the authors, $KLD(\pi(.|\xi) || \pi(.
|\xi_0))$ as opposed to $KLD(\pi(.|\xi_0) || \pi(. |\xi))$. In terms of interpretation, it is not obvious to us that one
conveys a better notion of complexity than the other.  However, looking at the various examples in the paper, we realised that had the
second choice been made the scaling argument in Section 3.3 would have failed and the PC prior would then be
undefined.  Indeed, in this case, the Kullback-Leibler divergence between the distributions $\mathcal N(0,\epsilon^2)$ and $\mathcal
N(0,\sigma^2)$ is given by $$ \frac{1}{2} \left( \frac{ \epsilon^2}{\sigma^2} - 1 - \log\left( \frac{
\epsilon^2}{\sigma^2}\right)\right) \approx \log\left( \frac{ \epsilon}{\sigma}\right) $$ when $\epsilon \approx 0$. Is
there an explanation as to why one divergence is better than the other that is
more illuminating than the mere observation than in the Gaussian case one works and the other fails?  

We furthermore like Eqn (3.1) as this equation shows how the base constraint takes one away from Jeffreys priors. 
 However, Eqn (3.1) does not seem
correct outside the unidimensional and bijective case, due to the differentiation term. Omitting this undefined Jacobian, one can opt
for a uniform prior on the Kullback spheres of radius $d(\xi)$. The main part of the paper conveys a feeling
of uni-dimensionality and we were eager to see how it extends to models with many hyperparameters, which happens in
Section 6. Similarly if $\xi $ where the mean of the Gaussian random variable and $\xi=0$ the base model, then $\xi \rightarrow d(\xi)$ is not bijective although symmetry arguments suggest that it is enough to define a prior on $|\xi|$.

There is also a potential difficulty in the feature that $d(\xi)$ cannot be computed in a general setting. (Assuming
that $d(\xi)$ has a non-vanishing Jacobian as on page 19 sounds rather unrealistic.) Still about Section 6, handling
reference priors on correlation matrices $\mathbf{R}$ is a major endeavour, which should produce a steady flow of
followers, even though it is certainly easier than contemplating the corresponding prior on a covariance matrix
\citep{barnard:mculloch:meng:2000}.

\begin{quote}``The current practice of prior specification is, to be honest, not in a good shape. While there has been a strong
growth of Bayesian analysis in science, the research field of ``practical prior specification'' has been left behind.''
(*p.23)\end{quote}

There are still (numerous) quantities to specify and calibrate in PC priors, which may actually be deemed a good thing by
Bayesians (and even by some modellers). But overall we think this paper and its central message constitute a terrific step for Bayesian
analysis and not solely for its foundations, provided a more directive approach is adopted.

\section{PC difficulties}

A first point in the delicate implementation of the PC principle is that those PC priors rely on several choices made in
the ordering of the relevance, complexity, nuisance level, and so on, of the parameters, in that way quite similar to reference priors
\citep{berger:bernardo:sun:2009}. While the first author also wrote a paper on Russian roulette \citep{lyne:2015}, there
is a further ``Russian doll" principle at work behind (or within) PC priors. Each shell of
the Russian doll corresponds to a further level of complexity whose order need be decided by the modeller. This does not
sound to be a very realistic assumption in a hierarchical model with several types of parameters having only local meaning.

A second point is that the construction of those PC priors reflects another Russian doll
structure, namely one of embedded models, hence would and should lead to a natural multiple testing methodology. Except
that the first author of the paper clearly rejected this notion during his talk at ISBA 2016, by being opposed to
testing {\em per se}. 

\section{Further remarks}

\begin{quote}"We do not know precisely the thinking that underscores the choice of prior, but we do know that they have
been hugely influential.  This is not a satisfactory state of affairs."\end{quote}

The paper repeats a well-known meme that computation killed the Bayesian spirit. Plus this less common notion that
priors should not be influential. Why not?! If there is no such thing as a non-informative prior, all priors are
influential and produce different inference outputs. What matters in the (user's) end is to provide a way to calibrate this output.

\begin{quote}"While reference priors have been successfully used for classical models, they
have a less triumphant history for hierarchical models."\end{quote}

This argument applies to any setting in the sense that reference priors do require some ordering in the importance or
relevance of the parameters. If there is subjectivity at this level, this also reflects on the ill-defined nature of
some components of a hierarchical model. Especially when considering models with latent variables.

\begin{quote}"To date, there has been no attempt to construct a general method for specifying weakly informative
priors."\end{quote}

We have no qualm about the section on {\em ad hoc} priors since we agree with the assessment that mimicking an earlier study
of a similar problem brings no justification to another use. And about the weakly informative section, where we agree
that the lack of general principles makes the argument hard to sustain.

The base model concept is once more appealing as a concept, reminding us of exponentially tilted models, although it is
 hard to see how to face the huge arbitrariness in setting base and extension. In that respect, Definition 1 is not
particularly helpful. And Informal Definition 1 makes things worse. Setting a base is clearly a subjective or personal
prior choice that should be acknowledged as such. About the desirable conditions, to which the authors are welcome and
free to set as they wish, {\bf D1} is not pertinent until the authors define the very notion of {\em non-informative}.
This of course sounds like a circular argument. Desideratum {\bf D2} makes sense only when provided the sampling model allows for a division
of the parameters into blocks. And {\bf D3} is also worth considering, if pretty vague. Desideratum {\bf D4} should first specify what
an {\em over-parameterised} model is, while {\bf D5} assumes identifiability is itself identifiable, which is not always the
case (although it may be the case for additive models). Desiderata {\bf D6} and {\bf D7} seem to proceed from common
sense, while {\bf D8} is not especially constrictive. Overall, we find it fairly hard to build a theory around such
vague principles, as they are too far from methodology, which brings back the earlier comments about the very notion of
Occam's razor. 

The principles set in Section 3 are making perfect sense, although we stress again that they do require a fair amount of calibration. And we
feel that the debate about using Cauchy versus Student's $t$ priors sounds like a bit of an old saw.  

\begin{quote}"PC priors are not built to be hypothesis testing priors and we do not recommend their direct use as
such."\end{quote}

There still is this feeling of a hypothesis test reformulation that occurs when considering the base model like the
null. See also the link with Johnson and Rossell (2010) non-local priors. We forgot about Bernardo's (2011) reformulation, which is under-exploited.

\begin{quote}"PC priors are defined on individual components. This distinguishes PC priors, from reference priors, in
which the priors depend on the global model structure.  This global dependence is required to ensure a proper posterior.
However, the modern applied Bayesian is far more likely to approach their modelling using a component-wise and often
additive approach."\end{quote}

Without here attempting a defence of reference priors \citep{bernardo:smith:1994} or of other non-informative construction techniques, we have
issues with this separation of the parameter in independent components. This feature remains an assumption and as such it may be a
poor choice. It is also hard to think of components of the parameter as being meaningful by themselves and thus to
contemplate building extended priors on some components irrespective of what happens to other parts of the parameter
somewhat defies reason. The remark on the PC prior construction in the context of sparse model presented above rules out to some extent the independence construction advocated by the authors. 

\begin{quote}" Close to the base model, the PC prior is a tilted Jeffreys' prior for $\pi(x|\xi)$, where the amount of tilting is determined by the distance on the Riemannian manifold to the base model scaled by the parameter $\lambda$."\end{quote}

It is not clear to us that the approximation proposed 
\begin{equation} \label{local} 
\pi(\xi)  = I(\xi)^{1/2} \exp(-\lambda m(\xi) ) + \cdots 
\end{equation}
is actually sharper than 
\begin{equation} \label{local2}
\pi(\xi)  = I(0)^{1/2}\lambda \exp(-\lambda \sqrt{I(0)} \xi ) + \cdots  
\end{equation}
which is a direct consequence of the first approximation provided by the authors 
$$
KLD(\pi(x|\xi) || \pi(x| \xi=0)) = \frac{ I(0)\xi^2}{2 } + \cdots
$$
Then locally the PC prior behaves like an exponential prior with parameter $\lambda \sqrt{I(0)}$. This representation is
less sophisticated than the authors' presentation, but it abstains from conveying the wrong notion that it locally resembles Jeffreys
prior. Indeed, it seems to us that the PC prior shares no common feature with Jeffreys priors, neither locally nor globally.

\section{Conclusion}

\begin{quote}"We still have to work them out on a case by case basis."\end{quote}

While the authors have uncovered several interesting and new avenues for exploring prior specification, we want to
signal yet another avenue, associated with the notion of Bayesian robustness, as proposed by \cite{holmes:watson:2016}
that we recently discussed in this journal (Robert and Rousseau, 2016). 

In fine, we congratulate the authors for this radical proposal that has the merit of defining a natural collection of
priors, while integrating the constraints of prior robustness. Formalising this most important aspect of Bayesian
modelling is absolutely essential for methodological and practical purposes, even if the current proposal is unlikely to
reach most practitioners.  We do acknowledge that the proposals made in the paper are currently exploratory, rather than
directive.  Again, we stress that this proposal represents an important step in the rational (if not objective)
construction of reference or weakly informative priors \citep{bad}.

\section*{Acknowledgements}
A component of those comments appeared on the first author's blog in a preliminary format. Discussions with Dan Simpson
and Havard Rue over several versions of the paper were most illuminating, starting at a 2014 workshop in Banff and continuing in
Warwick and Valenci{\`a} for O-Bayes 2015. We are grateful to James Watson for sharing with us his impressions on the
paper.  We are also grateful to Peter Green for his out-worldly patience and to the editorial team for helpful
suggestions towards the final version of this discussion.

\input{Simpsonetal.bbl}

\end{document}

%% file: Simpsonetal.bbl
\hyphenation{Post-Script Sprin-ger}

%% file: Simpsonetal.bbl
\begin{thebibliography}{14}
\expandafter\ifx\csname natexlab\endcsname\relax\def\natexlab#1{#1}\fi
\expandafter\ifx\csname url\endcsname\relax
  \def\url#1{\texttt{#1}}\fi
\expandafter\ifx\csname urlprefix\endcsname\relax\def\urlprefix{URL }\fi
\providecommand{\eprint}[2][]{\url{#2}}

\bibitem[{Barnard et~al.(2000)Barnard, McCulloch and
  Meng}]{barnard:mculloch:meng:2000}
\textsc{Barnard, J.}, \textsc{McCulloch, R.} and \textsc{Meng, X.-L.} (2000).
\newblock Modeling covariance matrices in terms of standard deviations and
  correlations, with application to shrinkage.
\newblock \textit{Statistica Sinica}, \textbf{10} 1281--1311.

\bibitem[{Berger et~al.(2009)Berger, Bernardo and
  D.}]{berger:bernardo:sun:2009}
\textsc{Berger, J.}, \textsc{Bernardo, J.} and \textsc{D., S.} (2009).
\newblock Natural induction: An objective {B}ayesian approach.
\newblock \textit{Rev. Acad. Sci. Madrid}, \textbf{A 103} 125--159.
\newblock (With discussion).

\bibitem[{Bernardo and Smith(1994)}]{bernardo:smith:1994}
\textsc{Bernardo, J.} and \textsc{Smith, A.} (1994).
\newblock \textit{{B}ayesian Theory}.
\newblock John Wiley, New York.

\bibitem[{Bochkina and Green(2014)}]{bochkina:green:2014}
\textsc{Bochkina, N.~A.} and \textsc{Green, P.~J.} (2014).
\newblock The {B}ernstein--von {M}ises theorem and nonregular models.
\newblock \textit{Ann. Statist.}, \textbf{42} 1850--1878.

\bibitem[{Gelman et~al.(2013)Gelman, Carlin, Stern, Dunson, Vehtari and
  Rubin}]{bad}
\textsc{Gelman, A.}, \textsc{Carlin, J.}, \textsc{Stern, H.}, \textsc{Dunson,
  D.}, \textsc{Vehtari, A.} and \textsc{Rubin, D.} (2013).
\newblock \textit{{B}ayesian {D}ata {A}nalysis}.
\newblock CRC press, Chapman \& Hall.

\bibitem[{Holmes and Watson(2016)}]{holmes:watson:2016}
\textsc{Holmes, C.} and \textsc{Watson, J.} (2016).
\newblock Approximate models and robust decisions.
\newblock \textit{Statis. Science} 27 p.
\newblock (To appear.).

\bibitem[{Johnson and Rossell(2010)}]{johnson:rossell:2010}
\textsc{Johnson, V.} and \textsc{Rossell, D.} (2010).
\newblock On the use of non-local prior densities in bayesian hypothesis tests.
\newblock \textit{J. Royal Statist. Society Series B}, \textbf{72} 143--170.

\bibitem[{Kass and Wasserman(1996)}]{kass:wasserman:1996}
\textsc{Kass, R.} and \textsc{Wasserman, L.} (1996).
\newblock Formal rules of selecting prior distributions: a review and annotated
  bibliography.
\newblock \textit{J. American Statist. Assoc.}, \textbf{91} 1343--1370.

\bibitem[{Liseo(2006)}]{liseo:2006}
\textsc{Liseo, B.} (2006).
\newblock The elimination of nuisance parameters.
\newblock In \textit{Handbook of Statistics} (D.~Dey and C.~Rao, eds.),
  vol.~25, chap.~7. Elsevier-Sciences.

\bibitem[{Lyne et~al.(2015)Lyne, Girolami, Atchadé, Strathmann and
  Simpson}]{lyne:2015}
\textsc{Lyne, A.-M.}, \textsc{Girolami, M.}, \textsc{Atchadé, Y.},
  \textsc{Strathmann, H.} and \textsc{Simpson, D.} (2015).
\newblock On {R}ussian roulette estimates for {B}ayesian inference with
  doubly-intractable likelihoods.
\newblock \textit{Statist. Sci.}, \textbf{30} 443--467.

\bibitem[{Martin and Liu(2015)}]{martin:liu:2015}
\textsc{Martin, R.} and \textsc{Liu, C.} (2015).
\newblock Marginal inferential models: Prior-free probabilistic inference on
  interest parameters.
\newblock \textit{J. American Statist. Assoc.}, \textbf{110} 1621--1631.

\bibitem[{Robert(2001)}]{robert:2001}
\textsc{Robert, C.} (2001).
\newblock \textit{The {B}ayesian Choice}.
\newblock 2nd ed. Springer-Verlag, New York.

\bibitem[{Seaman et~al.(2012)Seaman, Seaman and Stamey}]{hd}
\textsc{Seaman, J.}, \textsc{Seaman, J.} and \textsc{Stamey, J.} (2012).
\newblock {H}idden {D}angers of {S}pecifying {N}oninformative {P}riors.
\newblock \textit{American Statistician}, \textbf{66} 77--84.

\bibitem[{Zabell(1992)}]{zabell:1992}
\textsc{Zabell, S.} (1992).
\newblock Fisher and the fiducial argument.
\newblock \textit{Statist. Science}, \textbf{7} 369--387.

\end{thebibliography}
